\documentclass[aps,prl,showpacs,twocolumn,showkeys,amssymb,amsmath,superscriptaddress,reprint]{revtex4-1}

\usepackage{amsmath}
\usepackage{graphicx}
\usepackage{color}
\usepackage{tikz}
\usepackage{verbatim}

\usepackage{amssymb}
\usepackage{bm}
\usepackage{sidecap}
\usepackage{caption}

\newcommand{\rom}[1]{\textup{\uppercase\expandafter{\romannumeral#1}}}

\begin{document}

\title{Many-body entanglement in a topological chiral ladder}

\author{Ritu Nehra}
\author{Devendra Singh Bhakuni}
\author{Suhas Gangadharaiah}
\author{Auditya Sharma}
\email{auditya@iiserb.ac.in}
\affiliation{Department of Physics, Indian Institute of Science Education and Research, Bhopal, India}


\begin{abstract}
We find that the topological phase transition in a chiral ladder is
characterized by dramatic signatures in many body entanglement entropy
between the legs, close to half-filling. The value of entanglement
entropy for various fillings close to half-filling is identical, at
the critical point, but splays out on either side, thus showing a
sharp signature at the transition point. A second signature is
provided by the change in entanglement entropy when a particle is
added (or subtracted) from half-filling which turns out to be exactly
$-\log{2}$ in the trivial phase, but zero in the topological phase. A
microscopic understanding of tendencies to form singlets along the
rungs in the trivial phase, and along the diagonals in the topological
phase, is afforded by a study of concurrence. At the topological phase
transition the magnitude of the derivative of the average concurrence
of all the rungs shows a sharp peak. Also, at the critical point, the
average concurrence is the same for various fillings close to
half-filling, but splays out on either side, just like entanglement
entropy.
\end{abstract}

\maketitle

Topological states of
matter~\cite{bernevig2013topological,asboth2016short} have been at the
centre of physics research in the last decade or so. One of the
reasons for excitement has been the apparent simplicity of the models
involved underneath which rich physics lies, and continues to be
unearthed. Topologically signifcant states are often accompanied by
the presence of `edge states' with metallic properties while the bulk
is gapped and insulating.  It has been long realized that entanglement
in the many-body ground state can be a useful diagnostic for topological order. While the scaling to leading order of
entanglement entropy is governed by the famous `area law' ~\cite{hastings2007area,eisert2010}, it is the
subleading part that is linked with topological order and has now come to be known as topological entanglement
entropy~\cite{kitaev2006topological,levin2006detecting,jiang2012identifying}. A finer tool,
namely the entanglement spectrum has also been
widely used~\cite{li2008entanglement,ryu2006entanglement,bray2009topological,flammia2009topological,thomale2010entanglement,pollmann2010entanglement,prodan2010entanglement}.

In this Letter, we point out that entanglement in the many body ground
state, when considered in a comparative study of various fillings
close to half-filling may show a dramatic signature at a topological
phase transition. We choose a specific system, namely a two leg chiral
ladder, that has received a lot of attention in recent
times~\citep{hugel2014chiral,atala2014observation,zheng2017chiral},
but not from an entanglement perspective. This simple system turns
out to be rich with a Meissner to vortex phase transition, and in the
presence of diagonal hopping, a trivial-to-topological phase
transition. The study of entanglement in the many body and
single particle ground states in this system offers useful fresh
insights for not only the topological phase transition, but for the
Meissner to vortex phase transition as well.  In order to compute
entanglement entropy in the many body ground states of these systems,
we exploit the clever techniques of Peschel and
co-workers~\citep{peschel2012special,chen2013quantum}.

A further feature in our work is an investigation into the role of
concurrence~\cite{hill1997entanglement,wootters1998entanglement}, a
measure of two-site entanglement, whose study allows us to track
microscopic quantum correlations, and how they alter at the
topological phase transition. Any study of entanglement involves the
specification of subsystem and its complement; in the ladder system, a natural subsystem to work
with is one of the legs. With such a choice, it is of interest to understand not only the 
entanglement content between the two full legs, but also to consider the entanglement in each rung
separately, where concurrence proves to be handy.

\begin{figure}[h!]
\includegraphics[scale=0.85]{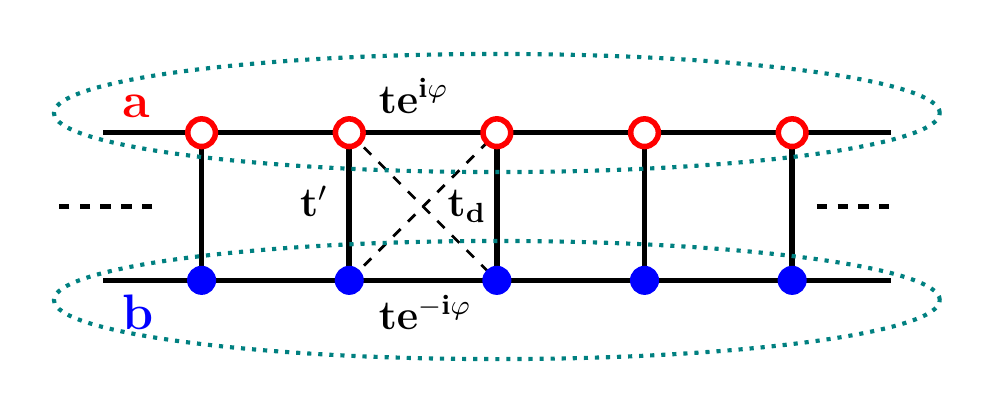}
\caption{The ladder system consists of two legs $\mathbf{a}$ and
  $\mathbf{b}$ with uniform magnetic flux $\phi$ per plaquette. The
  horizontal division for the computation of entanglement entropy, is delineated.}
\label{fig:1}
\end{figure}
The system consists of a two-leg ladder of non-interacting fermions subjected to a uniform magnetic flux $\phi$ per plaquette - a schematic diagram is given in Fig. \ref{fig:1}. The Hamiltonian can be written as~\cite{hugel2014chiral,atala2014observation,orignac2001meissner} 
\begin{align}
\label{eq:1} H=-t\displaystyle\sum_{\ell}&(e^{i\varphi}a_{\ell+1}^{\dagger}a_{\ell}+e^{-i\varphi}b_{\ell+1}^{\dagger}b_{\ell})-t^\prime\displaystyle\sum_{\ell}a_{\ell}^{\dagger}b_{\ell}\nonumber\\&-t_d\displaystyle\sum_{\ell}(a^{\dagger}_{\ell}b_{\ell+1}+b^{\dagger}_{\ell}a_{\ell+1})+H.c,
\end{align}
where the operator $a_{\ell}(b_{\ell})$ is the annihilation operator
at site $ \ell$ in the right(left) leg of the ladder. The parameters
$t$, $t^\prime$ and $t_d$ are the hopping amplitudes along the legs of
the ladder, along the rungs of the ladder without magnetic field and
along the diagonals of each plaquette, respectively, and $L$ is the length of the ladder. 
It is useful to define $\xi=\frac{t^{\prime}}{2t} ,\; \xi_d=\frac{t_d}{t}$. 
Using an appropriate gauge, the magnetic field is absorbed into the
hopping term ($t\rightarrow t e^{i\varphi}, \varphi=\phi/2$) by
Peierls substitution.

\begin{figure}[h!]
 \includegraphics[scale=0.76]{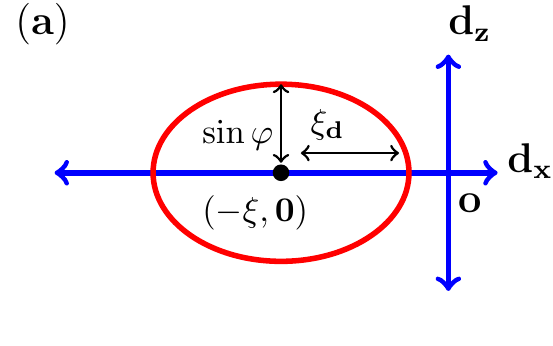}\
  \includegraphics[scale=0.76]{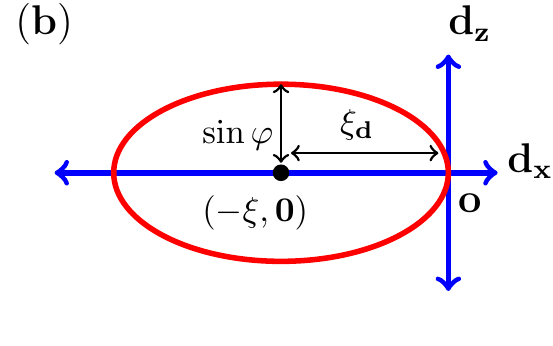}\
   \includegraphics[scale=0.85]{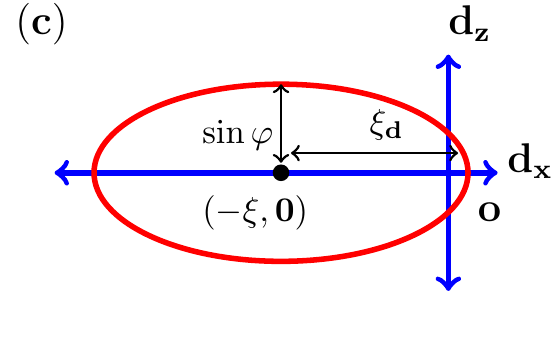}
\caption{The change in winding number with respect to change in $\xi_d$: (a) Trivial phase $\xi_d < \xi$, (b) the critical point $\xi_d = \xi$, and (c) the topological phase $\xi_d \geq \xi$. }
\label{fig:2}
\end{figure}

The ladder model shows a trivial-to-topological phase transition for
any general $\varphi$, which is signaled by a change in winding number. As shown in Fig.~\ref{fig:2}, there is
a change in winding number from $0$ to $1$ on increasing $\xi_d$
while keeping $\varphi$ constant. The Fourier transform of Eq.~\ref{eq:1} can be cast into the 
general form of a $2\times 2$ matrix in terms of Pauli matrices as $\mathcal{H}(k)=d_0 I +d_x\sigma_x +d_y\sigma_y+d_z\sigma_z$,
where $d_0=-\cos{\varphi}\cos k$, $d_x=-\xi-\xi_d\cos k$, $d_y=0$, and $d_z=-\sin{\varphi}\sin k$.
Here, only $d_x$ and $d_z$ contribute to the winding number calculation. Eliminating $k$, we have
the trajectory
\begin{equation}
\label{eq:2} \Big(\frac{d_x+\xi}{\xi_d}\Big)^2 + \Big(\frac{d_z}{\sin{\varphi}}\Big)^2 =1,
\end{equation}
from which the winding number around the origin is computed (Fig.~\ref{fig:2}).
So, there is a trivial-to-topological phase transition at $\xi=\xi_d$
for any $\varphi$; the only role of $\varphi$ is to alter the minor
axis length in the winding number calculation. 

The most widely used quantity to measure entanglement in a pure state
of a bipartite system is entanglement entropy, which is nothing but
the von Neumann entropy of the subsystem.  In general, this involves
the computation of the reduced density matrix followed by
diagonalization - often a daunting task for many body states, since
the size of the reduced density matrix typically scales exponentially with
the system size. However, for many body eigenstates of \emph{quadratic fermionic Hamiltonians}, the correlation 
matrix approach developed by Peschel and co-workers~\cite{peschel2012special}, facilitates this computation by 
reducing the diagonalization problem to order of the system size. We
adopt this approach to study entanglement in the many-body ground
state of our system. We show that the
entanglement contained in the many body ground state, close to
half-filling reveals striking features for the topological phase
transition.  The overall correlation matrix for the ladder is given by
\begin{equation}
\label{eq:3}C_{2L\times2L} =\begin{bmatrix}
&\langle a_m^\dagger a_n\rangle &\langle a_m^\dagger b_n\rangle\\
&\langle b_m^\dagger a_n\rangle &\langle b_m^\dagger b_n\rangle
\end{bmatrix},
\end{equation} 
where $m,n=1,2,\cdots,L$. To compute entanglement, one first selectively pulls out the part of the
correlation matrix which relates to the subsystem of interest and
diagonalizes this subsystem correlation matrix. The
entanglement entropy of the subsystem with respect to its complement
is given in terms of the eigenvalues $C_i$ of the subsystem correlation matrix by
\begin{equation}
\label{eq:4}S=-\sum_i \big(C_i\ln C_i +(1-C_i)\ln(1-C_i)\big).
\end{equation}

\begin{figure}[h!]
\includegraphics[scale=0.34,angle=-90]{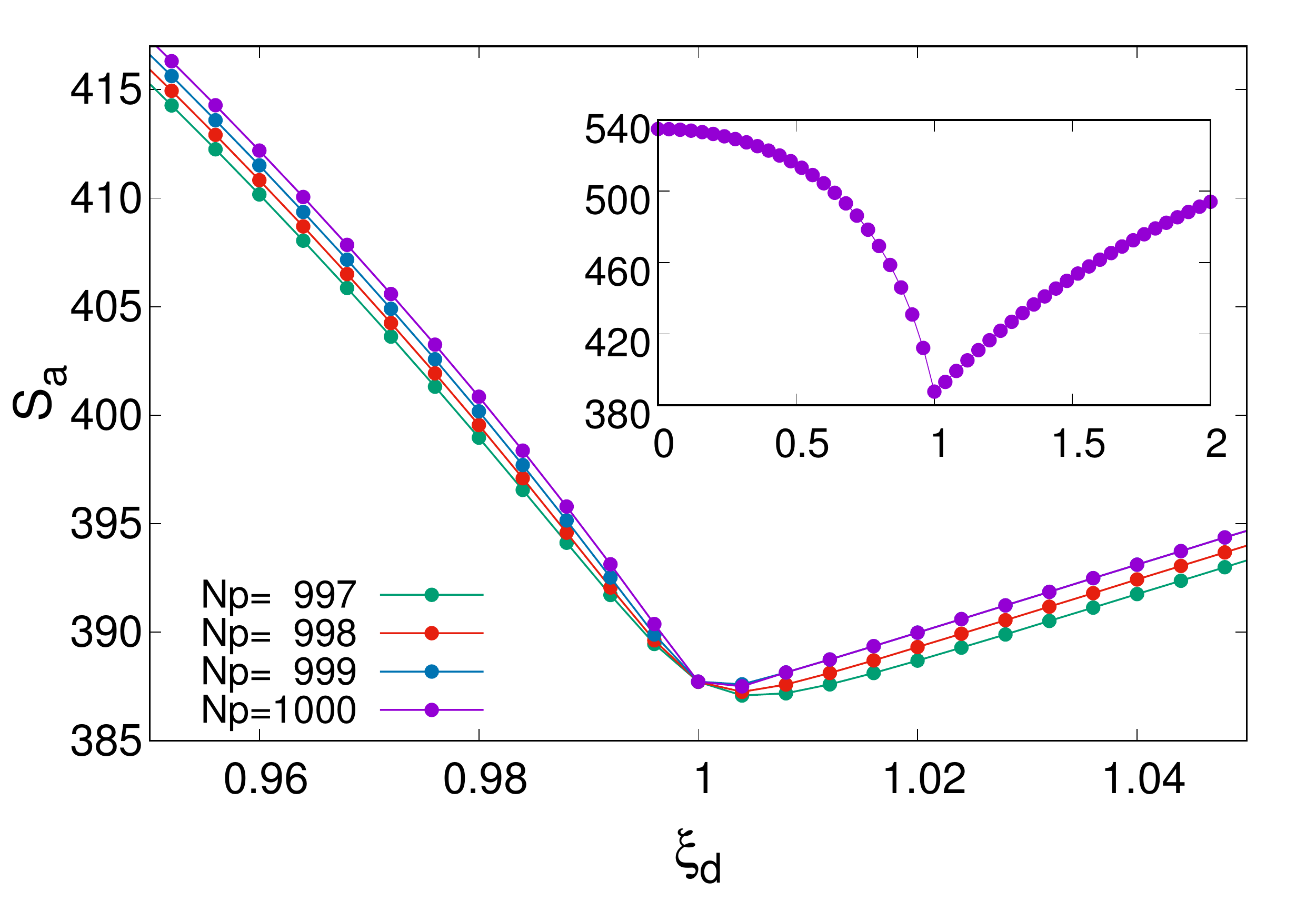}
\caption{The subsystem entanglement entropy ($S_a$) for the horizontal cut
  (Fig.~\ref{fig:1}) as a function of $\xi_d$, close to
  half-filling. Number of rungs $\text{L}=1000$,
  $\varphi=\frac{\pi}{2}$ and $\xi=1$ with open boundary
  conditions(OBC) imposed. $S_a$ takes on the
  same value for various fillings close to half-filling at the
  critical point. $S_a$ attains a minimum right after the topological phase
  transition $(\xi_d = 1)$. The inset shows only the data for
  half-filling in an extended region.}
\label{fig:3}
\end{figure}
Here we focus on the entanglement between the two legs of the ladder
as shown in Fig.~\ref{fig:1}. The ladder model lends itself naturally
to the horizontal division. Most studies of entanglement in chains (of the
SSH model~\cite{sirker2014boundary}, for example) have looked at the
vertical division, where a horizontal division does not exist. The
supplementary section does contain a discussion of entanglement in the
ladder model but with the other natural division, namely the vertical
division.  In the absence of diagonal hopping, when $\xi$ dominates,
the rungs of the system tend to form singlets. Furthermore, in the
limit where the legs hopping $t$ goes to zero, entanglement entropy
goes to $N_p\log2$, where $N_p$ is the number of particles, and
$\log2$ is the contribution from each singlet. As the legs hopping and
the magnetic flux contribution along the legs of the ladder are turned
on, a deviation from the value $N_p\log2$ is seen, although it
continues to be of this order of magnitude.  Fig.~\ref{fig:3} shows
the variation of entanglement entropy as a function of diagonal
hopping, in the vicinity of half-filling. In the trivial phase ($\xi_d
< \xi$), as the diagonal hopping is increased the singlets along the
rungs are systematically weakened, and therefore the entanglement
entropy decreases steadily. However, for $(\xi_d> \xi)$ edge states
appear, and form singlets (evidence for this comes from a study of
concurrence which appears later). This causes the entanglement entropy
to increase when $\xi_d$ is increased in the topological phase. In
addition, the large values of $\xi_d$ cause singlets to be formed
along the diagonals, which once again, in the limit of very large
$\xi_d$ yield a total entanglement entropy of $N_p\log2$, although
from a different mechanism here. The topological phase transition is
thus signalled by the entanglement entropy attaining a value
independent of filling in the vicinity of half-filling. The
entanglement entropy also attains a minimum soon after the topological
phase is entered; this minimum seems to be directly correlated with
the gap in the spectrum closing. 

\begin{figure}
\includegraphics[scale=0.3,angle=0]{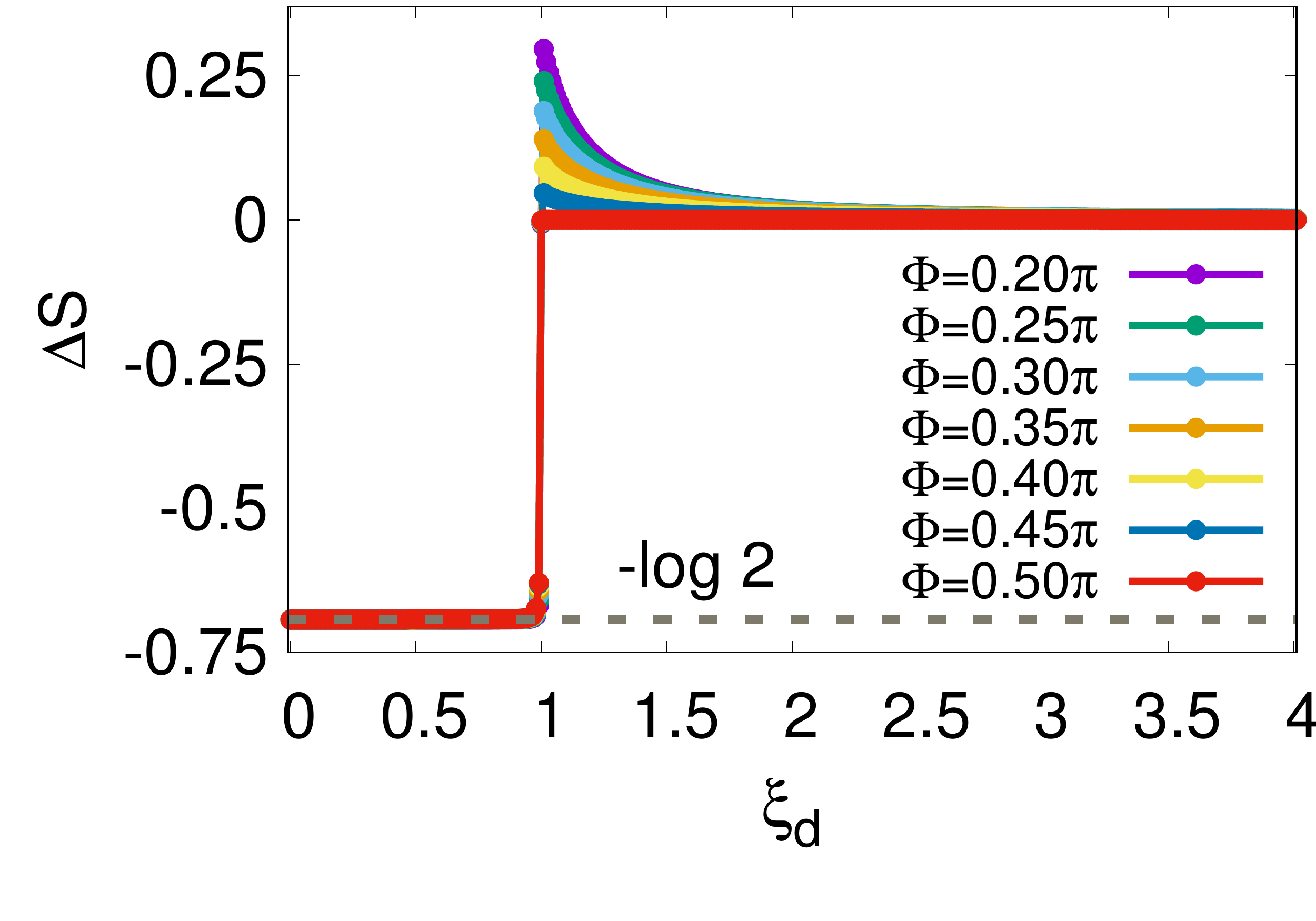}
\caption{$\Delta S$ as a function of $\xi_d$ under open boundary conditions (OBC) for various $\varphi$. $\xi=1.0$, $L=1000$.}
\label{fig:4}
\end{figure}

It is also insightful to study the entropy difference when a particle is either added or removed from
the half-filled state:
\begin{equation}
\label{eq:5}\Delta S = S_{\text{hf+1}}-S_{\text{hf}}.
\end{equation}
Fig.~\ref{fig:4} shows that in the topological phase $\Delta S$ goes
to zero, whereas in the trivial phase, $\Delta S$ is $-\log2$. The
$-\log2$ difference would be expected in the limit of the legs hopping
going to zero, because the half-filled state can then be thought of as
$N_p$ singlets. The removal or addition of one particle would then
result in the destruction of entropy equal to that of one
singlet. But, it is remarkable that this difference remains exactly
$-\log2$, even in the presence of legs hopping and flux.  In the
topological phase, the limit of large but finite $\xi_d$ is a useful
reference. At half-filling, exactly one of the edge states is
occupied, and this contributes zero to the entanglement, while the
remaining electrons form singlets along the diagonals. When one
particle is added to this state, it lands in the other edge, which
also contributes nothing to the entanglement, and thus $\Delta S$
would be zero. We see from Fig.~\ref{fig:4} that this feature is
exact throughout the topological phase for $\varphi = \pi/2$. For
other values of $\varphi$ though, we see that as one approaches the
critical point within the topological phase, the edge states do
contribute to the entanglement, thus causing $\Delta S$ to overshoot
zero. This seems to be related to the edge states being not completely localized at
the edges, when $\varphi$ is decreased.

In order to acquire a finer understanding of the nature of the many body ground
state wavefunctions, in various phases of the system, it is useful to study two-site entanglement.  An excellent
measure for this purpose is
concurrence~\cite{wootters1998entanglement,cho2017quantum,deng2004fermionic,zanardi2002fermionic,PhysRevA.93.032335,lakshminarayan2003entanglement}.
One nice feature of concurrence is that for a number
conserving Hamiltonian, the two-site concurrence is readily obtained,
regardless of whether the density matrix is pure or mixed. This
follows from the structure of the reduced density matrix for two sites
$i$ and $j$ $(\text{with} \ i<j)$, which can be written as
\begin{equation}
\label{eq:6}\rho_{ij} =\begin{bmatrix}
&u_{ij} &0 &0 &0\\
&0 &w_{1ij} &z_{ij}^{*} &0\\
&0 &z_{ij} &w_{2ij} &0\\
&0 &0 &0 &v_{ij}\\
\end{bmatrix},
\end{equation}
where, $u_{ij}=\langle(1-n_i)(1-n_j)\rangle,   w_{1ij}=\langle(1-n_i)n_j\rangle, w_{2ij}=\langle n_i(1-n_j)\rangle, v_{ij}=\langle n_{i}n_{j}\rangle \   \text{and}\  z_{ij}=\langle c_{j}^{\dagger}c_{i}\rangle$.
The concurrence is then given by 
\begin{equation}
\label{eq:7}\mathcal{C} = 2\;\text{max}(0,|z|-\sqrt{uv}).
\end{equation}
However, in the noninteracting framework, $\mathcal{C}$ can be
directly calculated from the subsystem correlation matrix of the two
sites. Employing Wick's theorem one can decompose the
four point correlators into two point correlators; the non zero
elements of the reduced density matrix $\rho_{ij}$ are then simplified
in terms of the correlation matrix.

\begin{figure*}
\includegraphics[scale=0.2325,angle=-90]{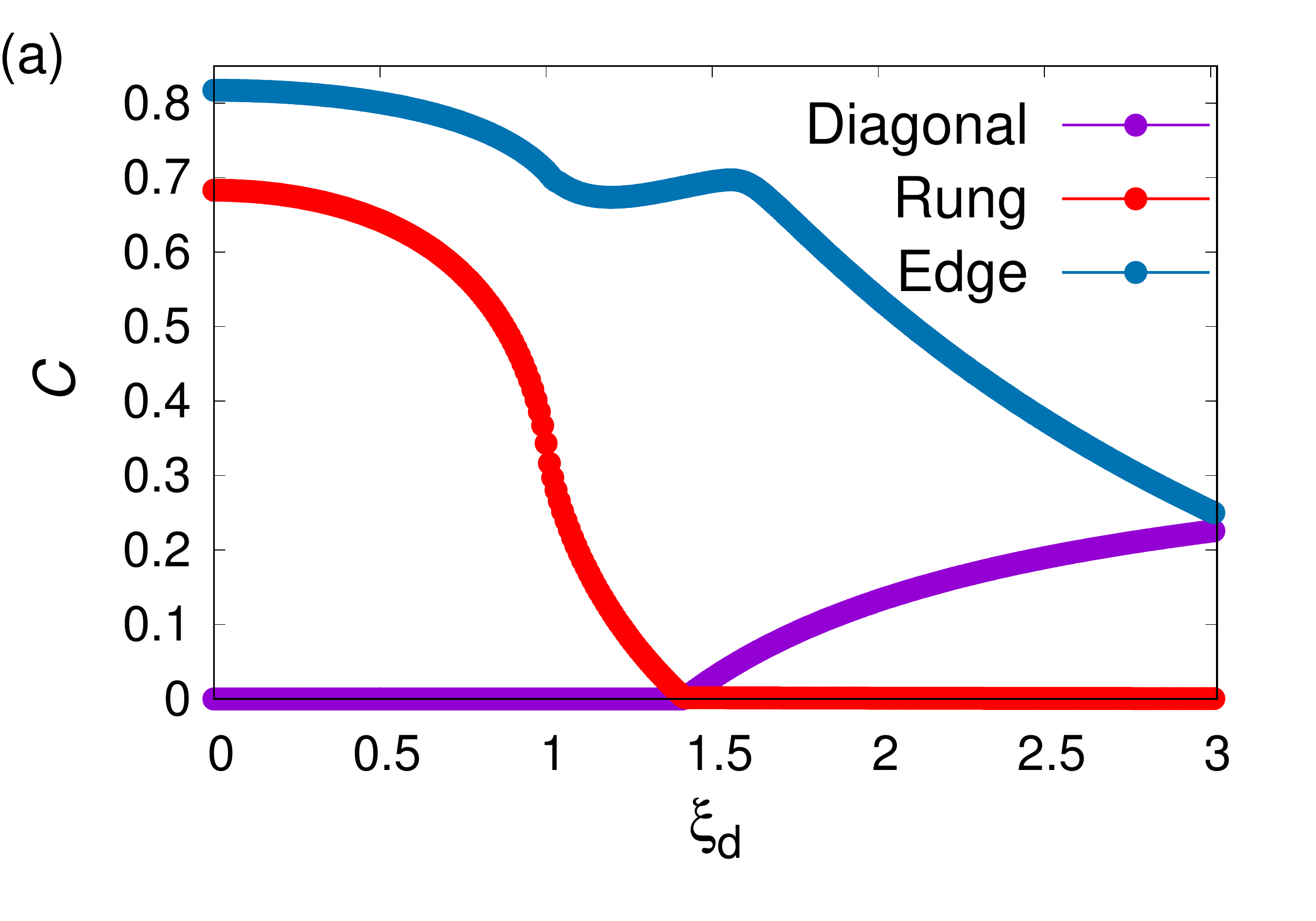}\
\includegraphics[scale=0.2325,angle=-90]{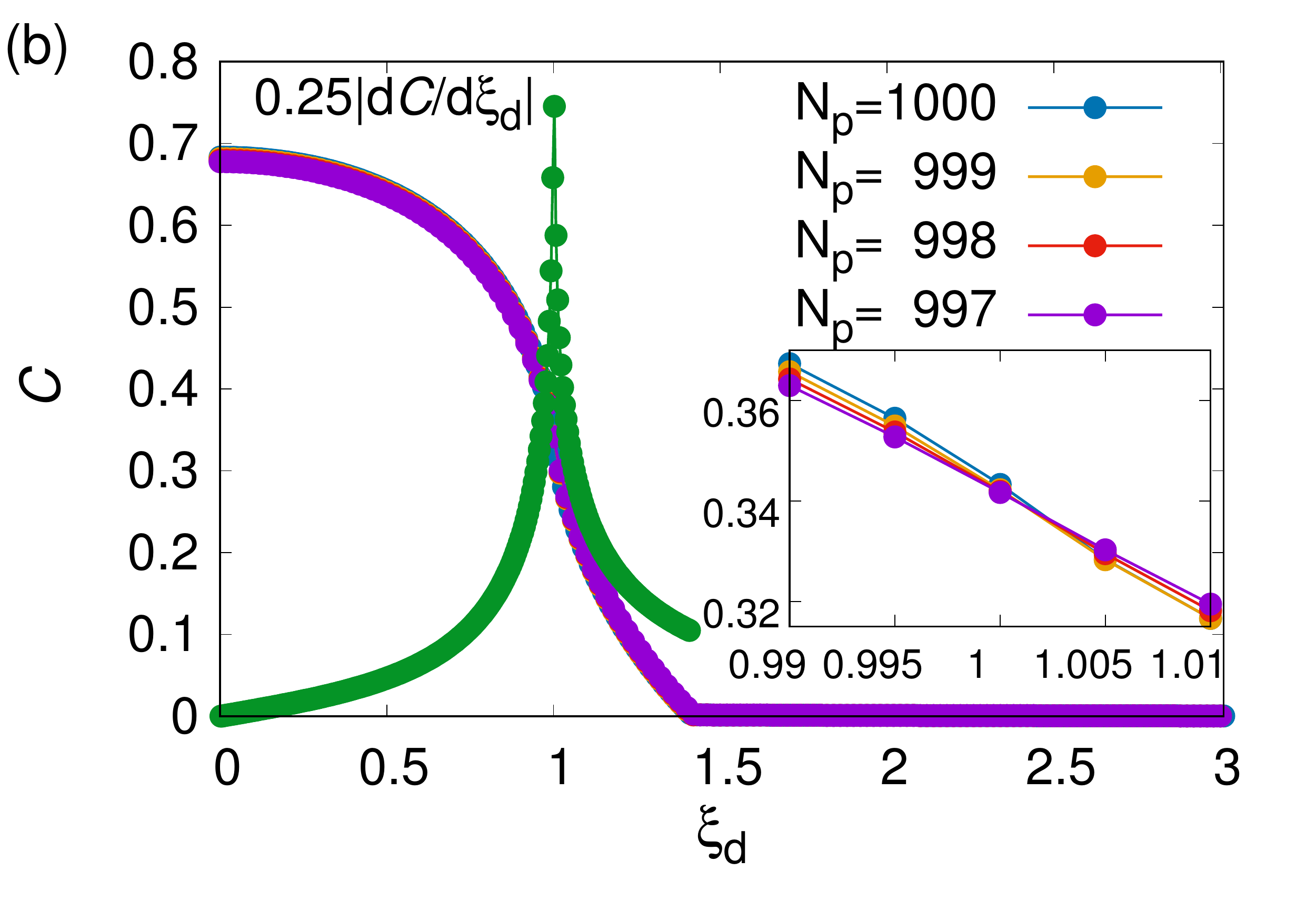}\
\includegraphics[scale=0.2325,angle=-90]{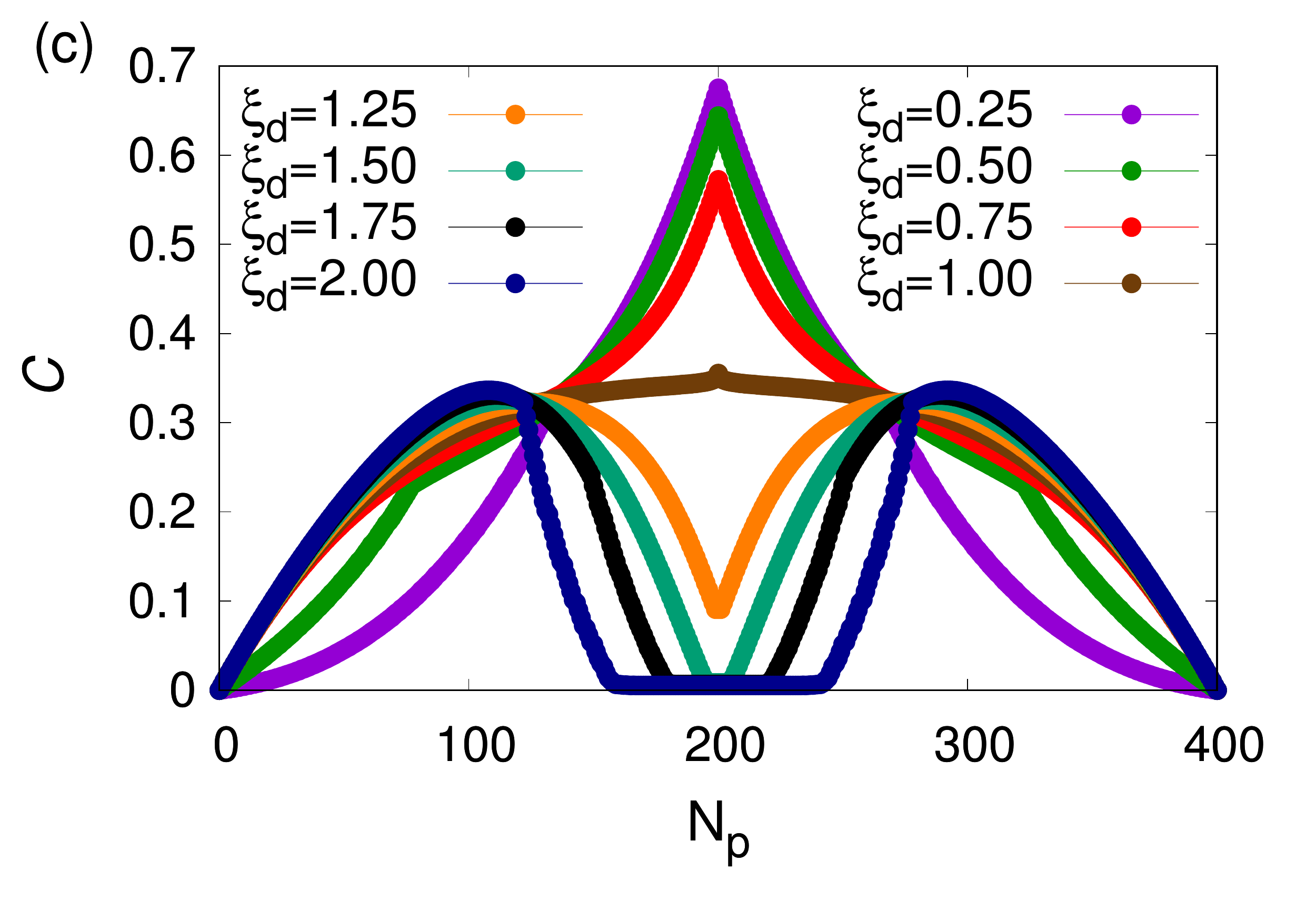}
\caption{(a) Average concurrence of diagonals, rungs and edge states
  of system with varying $\xi_d$ for half filling. (b) Average
  concurrence of rungs with $\xi_d$ near half filling. Also, the
  derivative of concurrence (green) featuring the
  trivial-to-topological transition. Inset shows a similar feature as Fig.~\ref{fig:3} for average rungs concurrence. (c) 
  Average concurrence of rungs as a function of filling, for various diagonal hopping. In all the figures, the parameters are $\xi=1.0$, $L=1000$ and $\varphi=\frac{\pi}{2}$ and
  open boundary conditions are imposed.}
\label{fig:5}
\end{figure*}

The concurrence bewteen two sites is maximum and equal to unity when
they form a singlet. For the ladder model, we can expect that the
system at half-filling, has a tendency to form singlets in each rung
when the rungs hopping is high and when the diagonal hopping is
small. As the diagonal hopping is increased, the tendency to form
singlets along the diagonals would be enhanced. The study of
concurrence reinforces these expectations. Fig.~\ref{fig:5}(a) shows the concurrence between the two
sites on a rung, averaged over all rungs, as a function of
$\xi_{d}$. Also included in the same figure is the concurrence between
the sites on a diagonal, averaged over all diagonals of the ladder. It
is seen that for large $\xi_d$, the rungs concurrence drops to zero,
whereas for small $\xi_d$, the diagonal concurrence is zero.
This type of a feature has been reported in the literature in the context of the
SSH model~\citep{cho2017quantum} and has been called `sudden-death of
concurrence' - it is a quantum information effect, and the value
of $\xi_d$ at which this change happens does not seem to have any
significance for the phases of the model. Moreover the $\xi_d$ values
at which the change happens for the rungs and the diagonal
concurrence, are close, but not identical.

A study of the concurrence between the edge sites is profitable for an
investigation into the role of quantum correlations within the edge states
in the topological phase. We notice that this shows a sharp change at
the topological phase transition point $\xi_d = \xi$. For $\xi_d \ge
\xi$, when the topological phase has just been entered, although the
overall concurrence between rungs continues to decrease, we observe
that the concurrence in the edge states increases in a brief
range. This suggests that the edge states when they have just formed
have singlet-like nature; however as the diagonal rungs are cranked
up, this character steadily decreases as the diagonals become more and
more singlet-like. The point at which the edge-state concurrence
begins once again to decrease seems to be connected to the appearance
of an enhanced density of states in the topological region, which is
discussed in the supplementary section.

It is also illuminating to study concurrence in the rungs, for a range
of fillings close to half-filling.  The inset of Fig.~\ref{fig:5}(b)
shows that at the topological phase transition, the average rungs
concurrence, similar to entanglement entropy, also attains the same
value for various fillings close to half-filling, and nicely splays
out on either side of the topological phase transition. The derivative
of concurrence shows a sharp feature at the topological phase
transition. Concurrence as a diagonastic at a phase transition has
been used in the SSH model~\citep{cho2017quantum}, the Harper model~\cite{lakshminarayan2003entanglement}, 
and in spin chains~\cite{osterloh2002scaling}.
Fig.~\ref{fig:5}(c) looks at the dependence of rungs concurrence, as a
function of filling, both in the trivial and the topological
phases. We see that in the trivial phase, concurrence has a peak at
half-filling, whereas this dramatically becomes a dip, as soon as the
topological phase is entered. Furthermore, in the limit of very large
diagonal hopping, this dip becomes a broad basin, close to
half-filling. At the topological phase transition, it is an almost
entirely smooth curve, except for a tiny peak at half-filling which
comes from the edge states - we have verified that the corresponding
model with periodic boundary conditions shows a completely smooth
curve, indicating that the tiny peak must indeed be a consequence of
the edge states.

To summarize, many body entanglement close to half-filling can provide
dramatic signatures at a topological phase transition. We show this by
considering the specific system of a chiral ladder. The entanglement
entropy between the legs of the ladder is independent of filling,
close to half-filling, if one is exactly at the topological critical
point, whereas this independence is lost on either side of the
transition. A similar feature is also shown by average concurrence in
all the rungs of the ladder. The magnitude of the derivative of this
concurrence has a dramatic peak at the transition point. Addition or subtraction of a particle
at half-filling can lead to either a precise change of $-\log{2}$ in entanglement entropy in the trivial
phase, or no change in the topological phase, due to the presence of edge states. In this
Letter, we have emphasized the usefulness of considering the
entanglement entropy between the two legs of the ladder in the
many-body eigenstates. A study of single-particle entanglement with
the same division captures the Meissner to vortex phase transition; these details can be found in the supplementary section. 
The study of entanglement entropy in the many-body ground state, but with a
vertical division of the subsystem provides further insights into the
topological phase transition - this too can be found in the
supplementary section. Our work opens up the question of how general these features of
many-body entanglement are for topological phase transitions. Recent work~\cite{zheng2017chiral} shows
that an electric field in this system can lead to chiral Bloch oscillations. Whether this can give rise
to special entanglement effects is worth investigating.

\acknowledgments A.S is grateful to SERB for the startup grant (File
Number: YSS/2015/001696).  \bibliography{ECL}
\end{document}


\title{Supplementary material for ``Many-body entanglement in a topological chiral ladder"}

\author{Ritu Nehra}
\author{Devendra Singh Bhakuni}
\author{Suhas Gangadharaiah}
\author{Auditya Sharma}
\affiliation{Department of Physics, Indian Institute of Science Education and Research, Bhopal, India}


\maketitle

\section{\label{sec:level2}Ladder Hamiltonian}
Since the Hamiltonian is translationally invariant along the legs, its momentum space version reads
\begin{equation}
H=2t\displaystyle\sum_{k}c_{k}^{\dagger}\mathcal{H}(k)c_{k},
\end{equation} 
where
 \begin{equation}
 \label{eqs:1}
\mathcal{H}(k)=
\begin{bmatrix}
-\cos(\varphi-k)& -\xi-\xi_d\cos k\\
-\xi-\xi_d\cos k& -\cos(\varphi+k)
\end{bmatrix},
\end{equation}
with 
$c_{k}^{\dagger}=\begin{bmatrix}
a_{k}^{\dagger}&b_{k}^{\dagger}
\end{bmatrix},\;c_{k}=\begin{bmatrix}
a_{k}\\b_{k}
\end{bmatrix},\; \xi=\frac{t^{\prime}}{2t} ,\; \xi_d=\frac{t_d}{t}$.
The dispersion consists of two bands with $E_+(E_-)$ being the energies of higher(lower) energy band:
\begin{equation}
E_{\pm}(k)=-\cos\varphi\cos k\pm\sqrt{(\xi+\xi_d\cos k)^2+\sin^2\varphi\sin^2k}.
\end{equation}
The corresponding eigenvectors are
\begin{align}
\label{eqs:2}
\Gamma^\dagger_{k}|0\rangle&=\frac{1}{N}\Big((\xi+\xi_d\cos k) a^\dagger_{k}+Y b^\dagger_{k}\Big)|0\rangle\\
\Omega^\dagger_{k}|0\rangle&=\frac{1}{N}\Big(-Y a^\dagger_{k}+(\xi+\xi_d\cos k) b^\dagger_{k}\Big)|0\rangle,
\end{align}
where $\Gamma^\dagger_{k}$,\;$\Omega^\dagger_{k}$ are new creation operators for the lower($E_-$) and higher($E_+$) energy bands respectively, with $Y=(\sqrt{(\xi+\xi_d\cos k)^2+\sin^2\varphi\sin^2k}\;-\;\sin\varphi\sin k)$ and $N=\sqrt{(\xi+\xi_d\cos k)^2+Y^2}$ is the normalization constant.
\section{\label{sec:level3} The vortex to Meissner phase transition}
Entanglement in the single-particle states provides useful signatures of the vortex-to-Meissner phase transition~\cite{hugel2014chiral,atala2014observation}, which will be described in this section.
\subsection{No diagonal hopping}
In the absence of diagonal hopping ($t_d=0$), there are two
parameters: one is the uniform magnetic field and the other is
$\xi$. For a constant $\xi$ and increasing $\varphi$, the dispersion curve
shows a phase transiton from the Meissner to the vortex phase at a certain
critical value of $\varphi$.
\begin{figure}[h!]
 \includegraphics[scale=0.72]{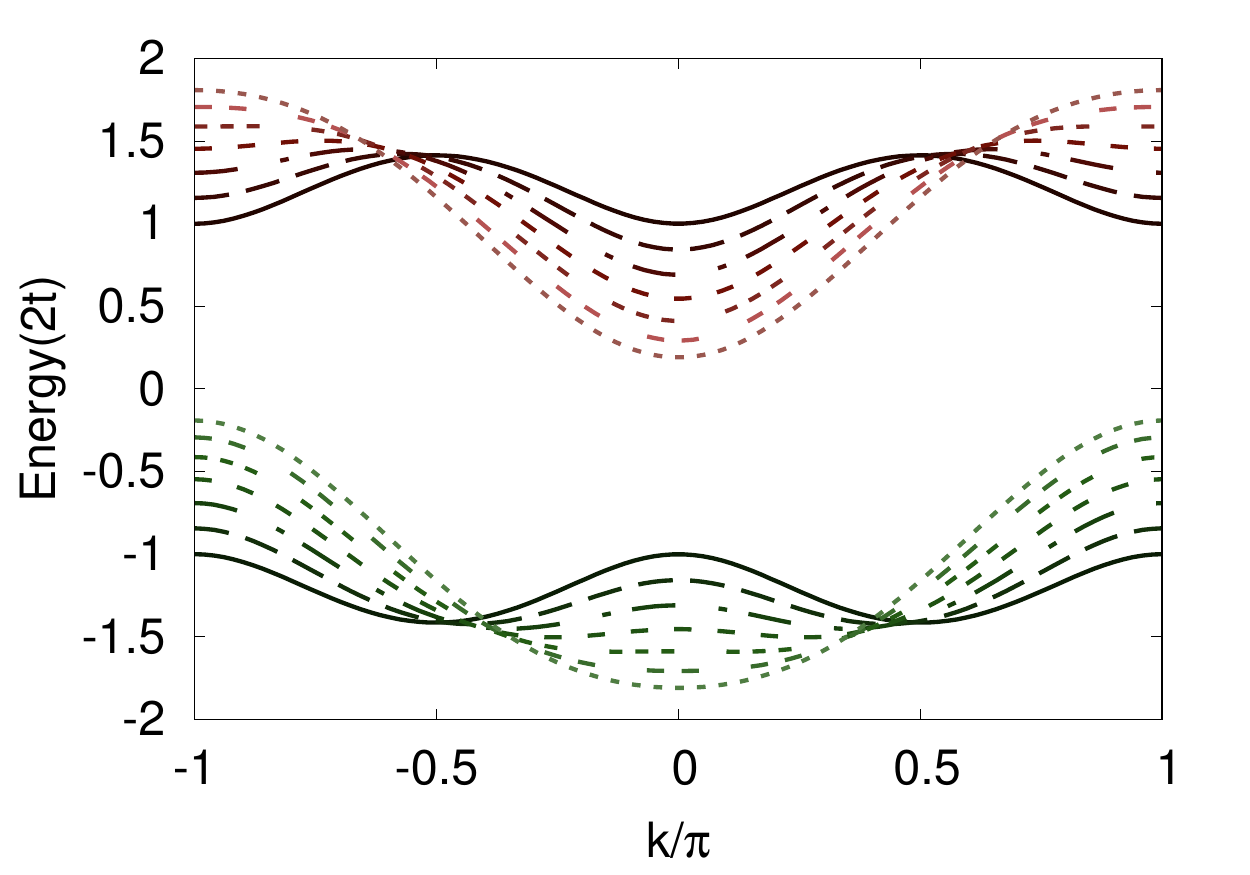}
\caption{ Lower energy band (green color) and higher energy band (red color) for the ladder system with the parameters $\xi=0.5,\;\xi_d=0$. Decreasing intensity of color (length of dots) corresponds to decreasing $\varphi$ starting from $0.5\pi$, in units of $0.05$ down to $0.2\pi$.}
\label{figs:1}
\end{figure}
From the dispersion curves in Fig. \ref{figs:1}, it can be seen that 
the lower energy band of the system possesses two minima at $\pm k_g$ which become one minimum at $k=0$, on decreasing $\varphi$ below a cricitcal value $\varphi_c$. The critical value $\varphi_c$ is obtained by minimizing the
lower band energy $E_{-}(k)$ with respect to $k$, and demanding that $k_{g}\neq 0$. This yields
\begin{equation}
\label{eqs:3}\varphi_c=\cos^{-1}\Big(\frac{-\xi+\sqrt{\xi^2+4}}{2}\Big)
\end{equation}
\begin{equation}
\sin k_g=\pm\sqrt{\sin^2\varphi_c-\cot^2\varphi_c\xi^2}.
\end{equation}
This phase transition is captured by other quantities like the chiral current~\cite{hugel2014chiral}. The currents in the two legs are 
\begin{align}\label{eqs:4}
J_a=\sum_l J_{a_{l}}=\frac{i}{2}\sum_l\langle e^{i\varphi}a_{l}^{\dagger}a_{l-1}-e^{-i\varphi}a_{l-1}^{\dagger}a_{l}\rangle\\
J_b=\sum_l J_{b_{l}}=\frac{i}{2}\sum_l\langle e^{-i\varphi}b_{l}^{\dagger}b_{l-1}-e^{i\varphi}b_{l-1}^{\dagger}b_{l}\rangle.
\end{align}
So, the chiral current is given by $J=J_a-J_b$.
In the single particle ground state when $\xi$ is kept constant, the chiral current
increases sinusoidally for increasing magnetic flux upto the
critical point ($\varphi_c$). On increasing $\varphi$ beyond the
critical point, it starts to decrease suggesting a phase transition
from Meissner phase to vortex phase as shown in Fig.~\ref{figs:2}(a).
\begin{figure}[h!]
\includegraphics[scale=0.56]{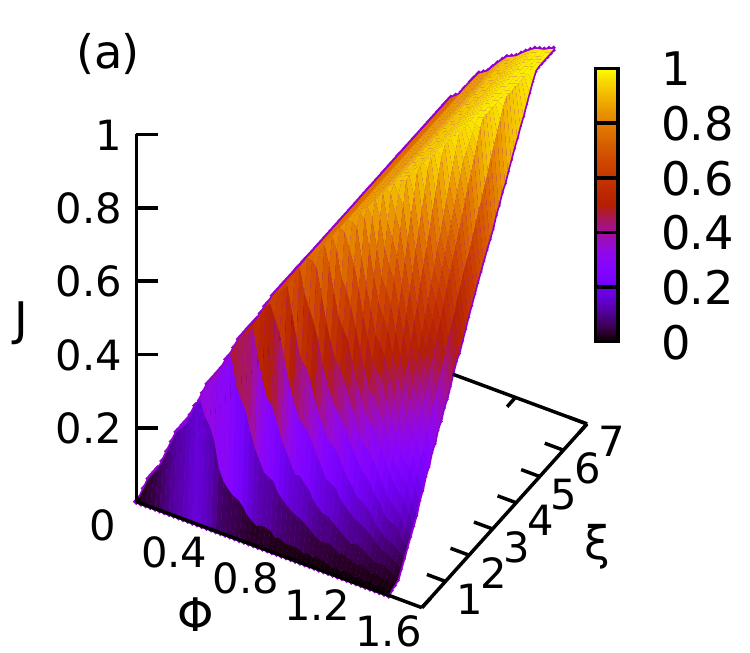}\
\includegraphics[scale=0.56]{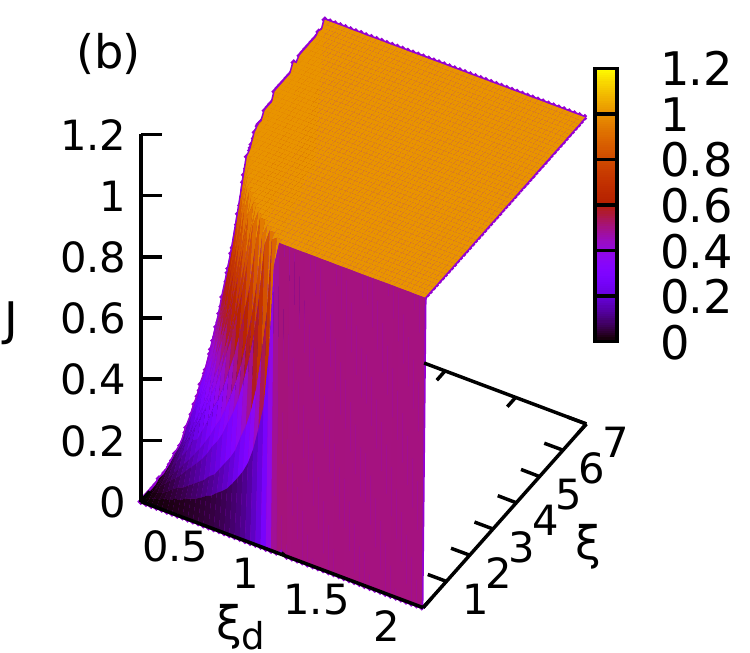}
\caption{Variation of chiral current ($\mathbf{J}$) for one particle ground state with parameters (a) $\xi$ and $\varphi$ with $\xi_d = 0$, and (b) $\xi$ and $\xi_d$ for a constant $\varphi=\frac{\pi}{2}$.}
\label{figs:2}
\end{figure}
The same kind of phase transition can be seen on increasing the
relative hopping strength ($\xi$) of the ladder for a constant
magnetic field, as was reported earlier~\citep{hugel2014chiral}. The
chiral current initially increases with increase in $\xi$ and beyond
the critical point ($\xi_c$) it saturates indicating the vortex-to-Meissner
phase transition.

\begin{figure}[h!]
\centering
\includegraphics[scale=0.56]{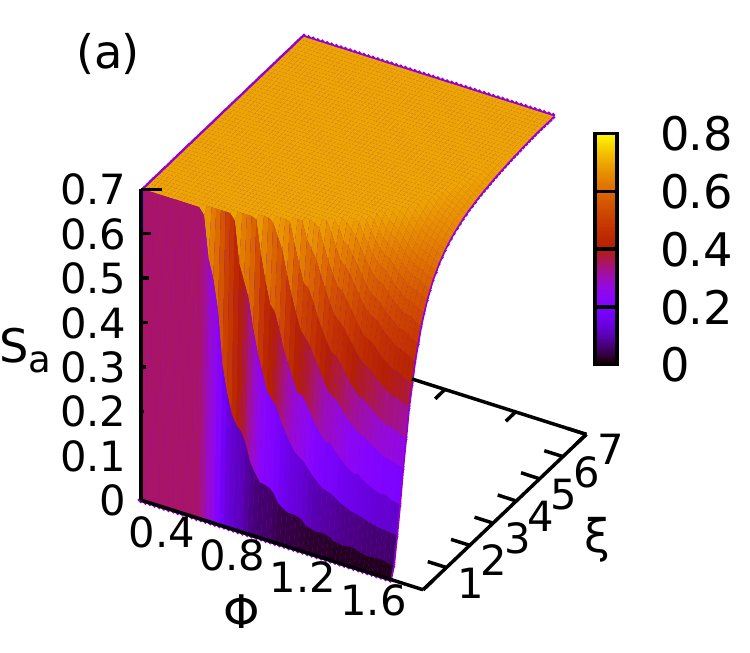}\
\includegraphics[scale=0.56]{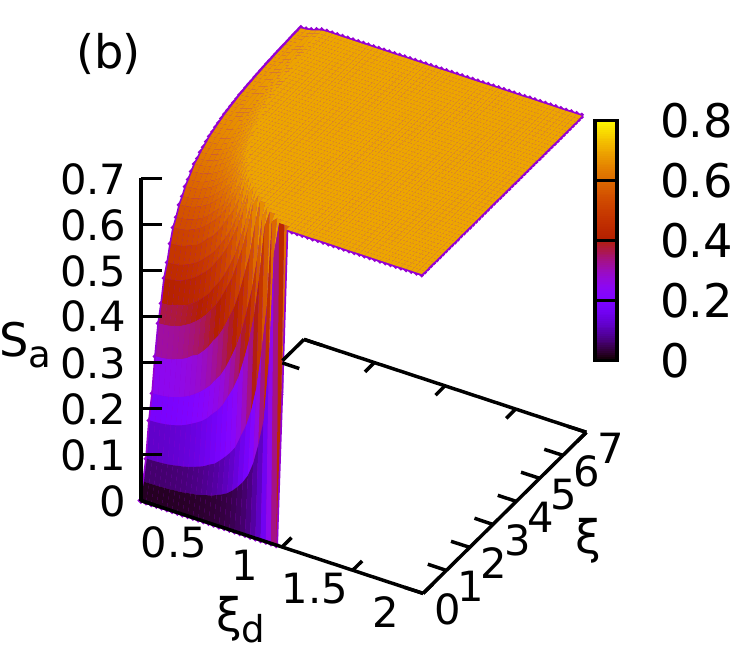}
\caption{Entropy of subsystem \textbf{a} ($S_a$) of one particle ground state as a function of (a) $\xi$ and $\varphi$  (b) $\xi$ and $\xi_d$ i.e. diagonal hopping and $\varphi=\frac{\pi}{2}$.}
\label{figs:3}
\end{figure}

Next we study the entanglement between the two legs of the ladder; in
other words, we consider the horizontal division as shown in Fig.1 of
the main paper. The vortex-to-Meissner transition is nicely captured
at the single particle level. The two-point correlations restricted to
the subsystem lattice sites yield the subsystem correlation matrix,
from which the entanglement entropy can be computed. The subsystem
correlation matrix pertaining to $\mathbf{a}$ is
\begin{equation}
\label{eqs:5}
C^{\mathbf{a}}=\Bigg[\frac{1}{L} \sum_k e^{-ik(m-n)}\frac{(\xi+\xi_d\cos k)^2}{N^2}\Bigg]_{L\times L},
\end{equation}
where the summation over $k$ is made over all the occupied levels. For single particle ground states, it would just be one number.
Fig.~\ref{figs:3} shows subsystem entropy behaviour with change in $\varphi$ and $\xi$. 

Here, for small $\varphi$ with constant $\xi$ the system is in the
maximally entangled state i.e. an entropy of $\log 2$. This is because
for small $\varphi$ the system has a unique minimum for the lower
energy band, which indicates that the particle could be on either of
the legs of the ladder with equal probability (Eq.(\ref{eqs:2}) with $\xi_d = 0$), which
in turn means that the entropy is maximal. On increasing $\varphi$
beyond $\varphi_c$ there are two degenerate minima for the lower
energy band i.e. the particle is either on the upper leg
($\mathbf{a}$) for positive momentum or on the lower leg
($\mathbf{b}$) for negative momentum. Since the lack of information is
minimal, this results in low entanglement entropy of the
subsystem. Similar behaviour is shown with variation of $\xi$ keeping
$\varphi$ constant. For $\xi=0$ the two legs are disconnected, and
therefore the entanglement entropy for subsystem $\mathbf{a}$ is
zero. It increases on increasing $\xi$ till $\xi_c$. Thereafter, it
saturates because the minimum is unique and at $k=0$ i.e. the
maximally entangled state.  

\subsection{\label{sec:level1} Diagonal hopping}
When diagonal hopping $t_{d}$ is turned on, the ladder system is rich
with multiple phases. First, there is still the vortex to Meissner
phase transition.  Once again, signatures for this transition are seen
both in chiral current shown in Fig.~\ref{figs:2}(b), and in
entanglement entropy for the same conditions as shown in
Fig.~\ref{figs:3}(b).  Diagonal hopping favours the Meissner phase as
indicated by Eq.~\ref{eqs:2}. It also contributes to making the
probability of the particle being in either leg equiprobable in the
Meissner phase, and thus making the ground state maximally entangled.
\begin{figure*}
\includegraphics[scale=0.165,angle=-90]{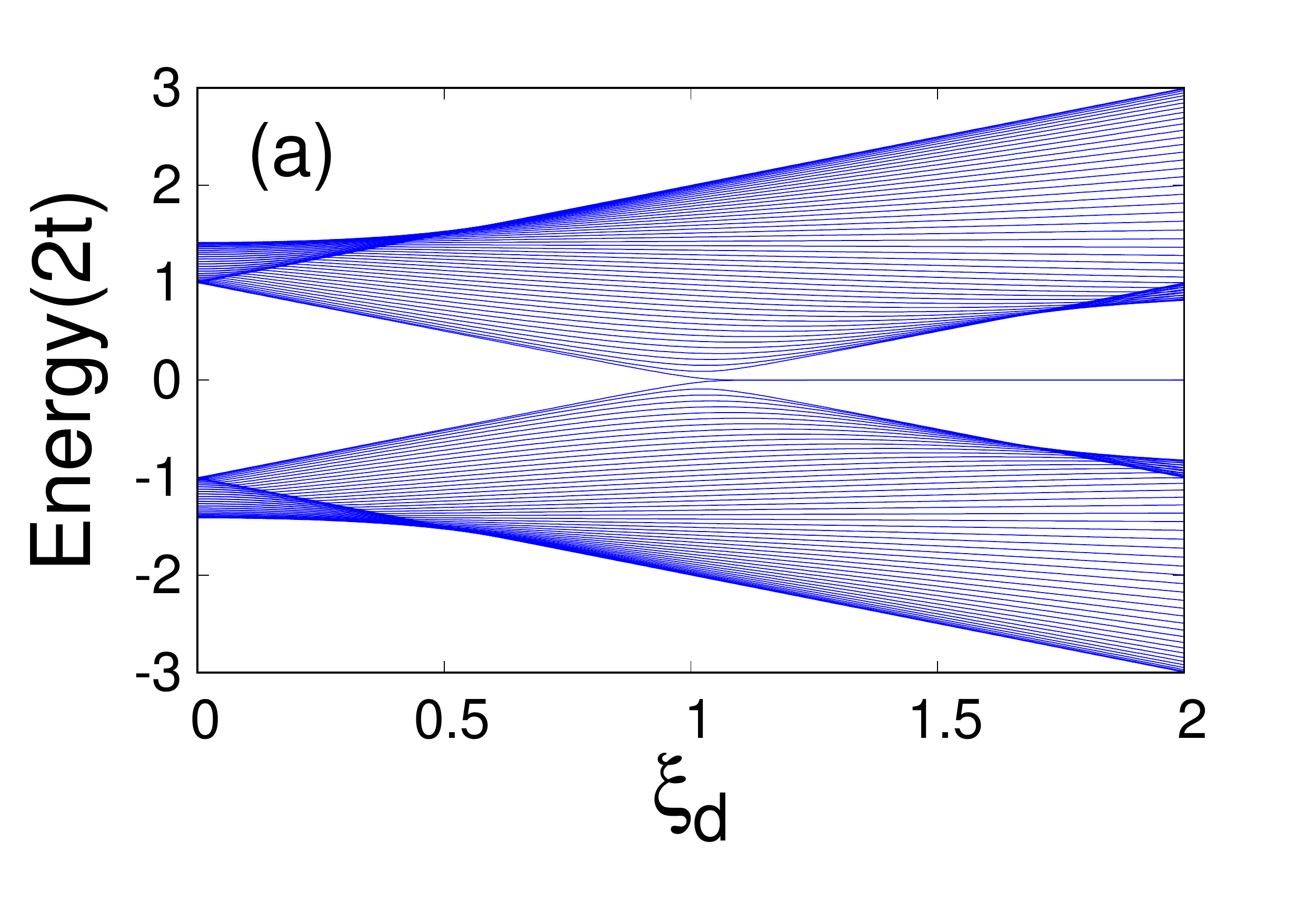}
\includegraphics[scale=0.165,angle=-90]{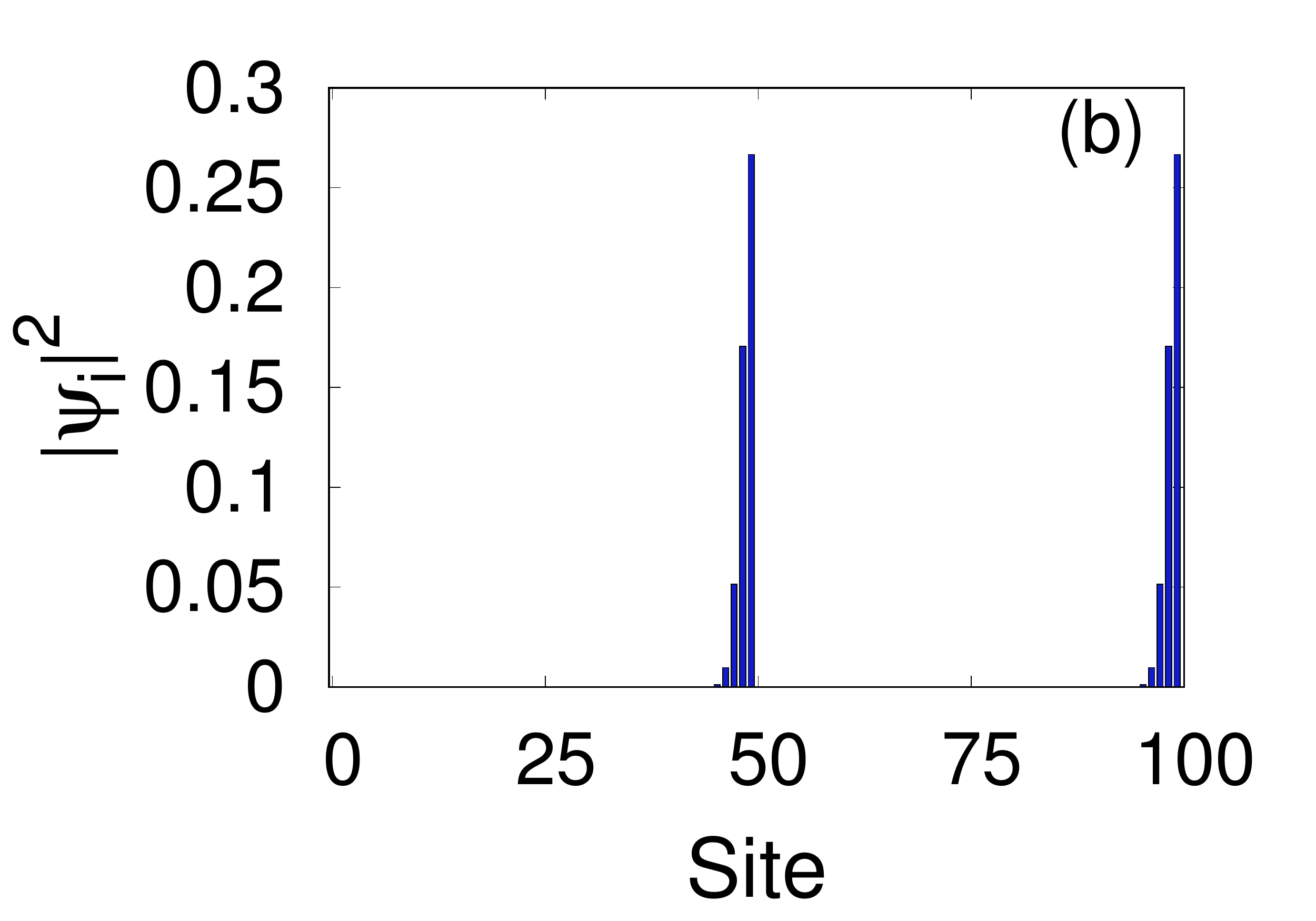}
\includegraphics[scale=0.165,angle=-90]{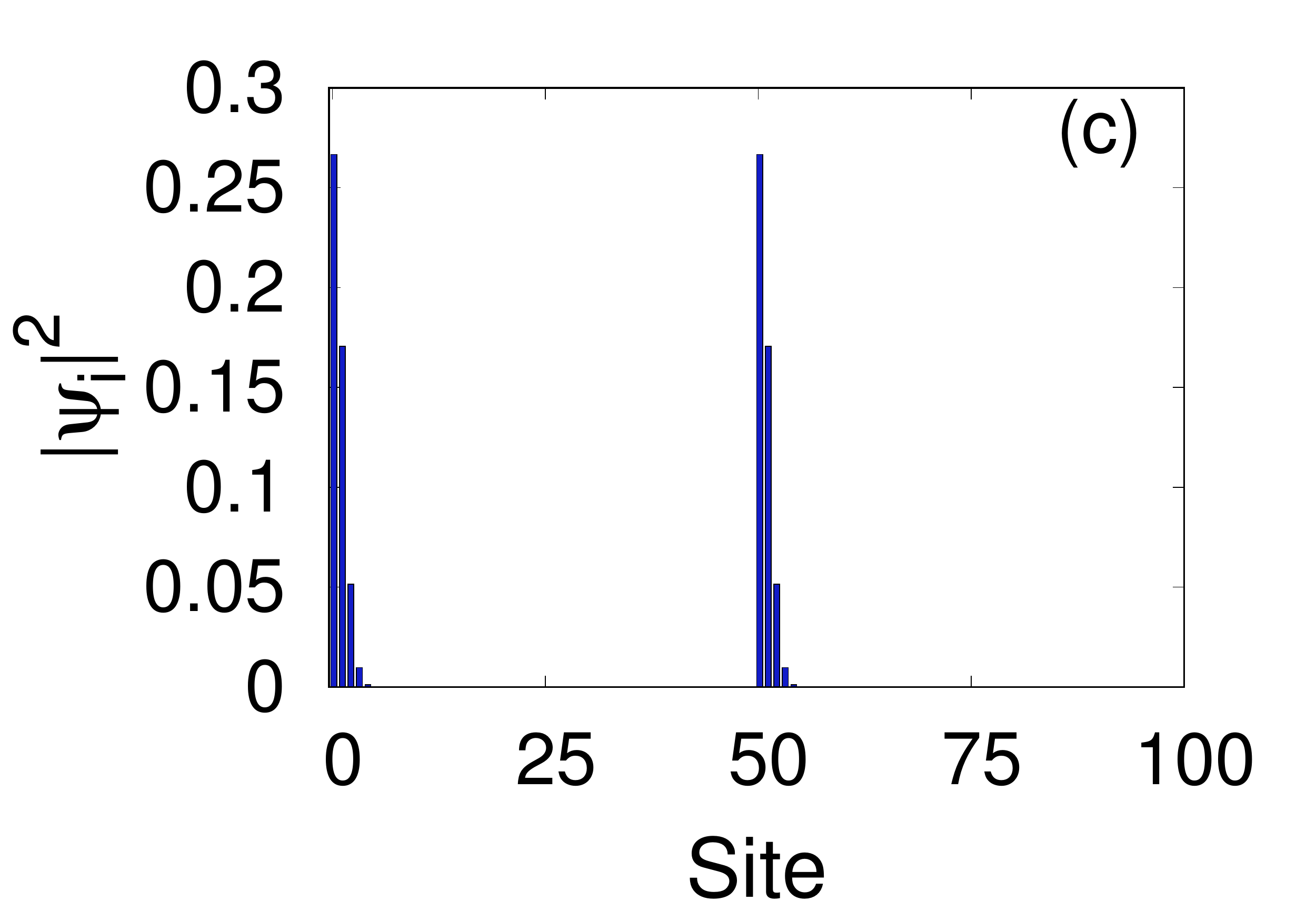}
\includegraphics[scale=0.165,angle=-90]{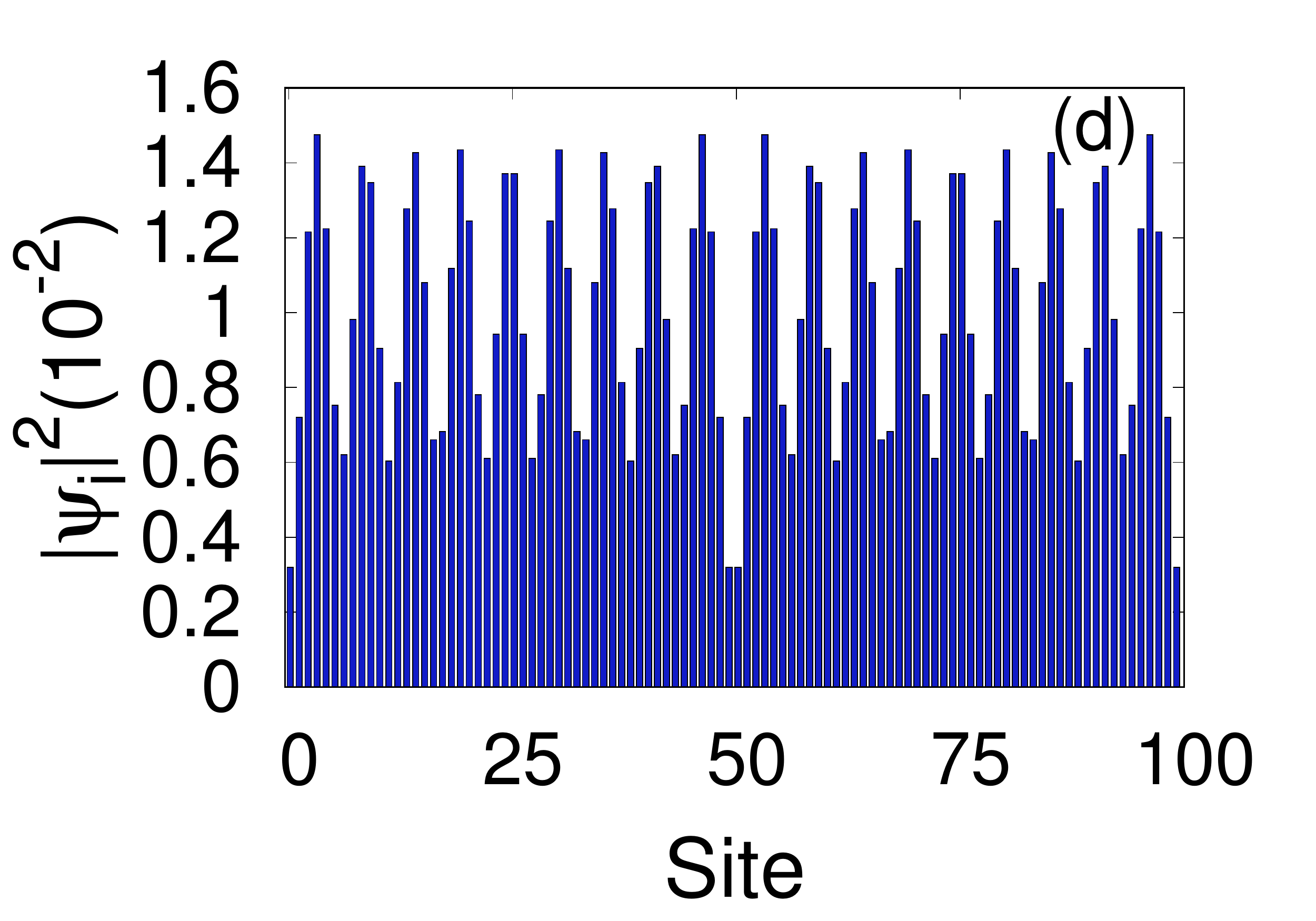}
\caption{(a) Energy spectra for ladder system with varying $\xi_d$.
  Squared amplitudes of the coefficients of
  wave function (${\psi_i}^2$) for edge sates are show in (b) and (c), and for a typical bulk state in (d), with $\xi_d$ fixed at $1.5$. 
  The other parameters $\xi=1$, number of rungs $L = 50$, $\varphi=\frac{\pi}{2}$ are common to all figures. 
  The indices first run through $1$ to $L$ among the `a' sites, and then again $1$ to $L$,
  through `b' sites, hence the edge-states show signals at one edge and close to the centre of the figure, which is also an edge of the ladder.}
\label{figs:4}
\end{figure*}

\begin{figure}
(a)\includegraphics[scale=0.7]{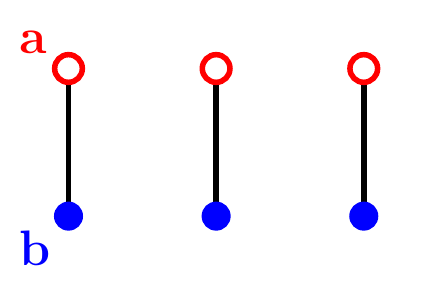}
(b)\includegraphics[scale=0.7]{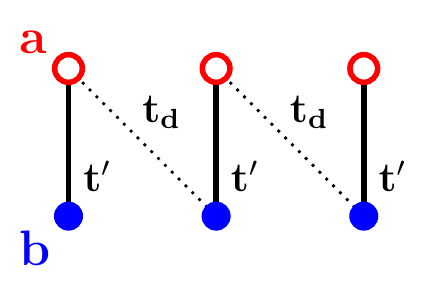}
\caption{Visualizing the SSH model as being made up of two-site unit cells.}
\label{figs:5}
\end{figure}

Diagonal hopping also induces a topological phase transition, which is
the focus of the main paper. With open boundary conditions (OBC), the
energy spectra for a constant $\xi$ and a constant magnetic flux with
varying $\xi_d$ is shown in Fig.~\ref{figs:4}(a). A pair of zero energy
states i.e. edge states appear for $\xi_d\ge\xi$ signalling a
trivial-to-topological phase transition in the system. We observe that
for $\varphi = \frac{\pi}{2}$, the spectrum is symmetric about the
zero energy modes. For other $\varphi$, the symmetry is
broken; however, the toplogical phase transition always happens at the
same point $\xi_d=\xi$, independent of $\varphi$. 
The case $\varphi=\frac{\pi}{2}$ is special. Here, the
unitary transformation $ U_c = (\sigma_z + \sigma_y )/\sqrt{2}$
applied to eq.~\ref{eqs:1} yields a structure similar to that of the
SSH model. One way to visualize the
ladder model as a generalization of the SSH model is as
follows. Consider a series of unit cells each consisting of two-sites
(one of type `a' and the other of type `b') as shown in Fig.~\ref{figs:5}(a). If the neigboring unit
cells are connected in such a way that only type `b' site of one unit
cell couples with type `a' of the neigbouring cell, then the SSH model
is obtained (Fig.~\ref{figs:5}(b)).  On the other hand, if every site of a unit cell
couples with every site of the neighbouring cell, then the ladder model
is obtained. The momentum space Hamiltonian under a suitable unitary
transformation yielding a structure identical to that of the SSH model
is a further consequence of the special case of $\varphi = \frac{\pi}{2}$.
In the SSH model, the wave functions for the two edge states are localized
on the ends of the chain.  Analogously, in the ladder model, edge states (Figs.~\ref{figs:4}(b) and ~\ref{figs:4}(c)) are localized on
the ends (first and last unit cells) i.e. ($a_1$, $b_1$) and
($a_L$, $b_L$) and decay exponentially in the bulk. Unlike edge states, any
other typical wave function of the system has some random distribution over all the
sites as shown in Fig.~\ref{figs:4}(d).

Apart from the topological phase transition $\xi_d$ can also trigger
the vortex to Meissner phase transition.  Below a critical value
$\xi_d\le{\xi_{d}}_{c}$ in the vortex phase~\cite{hugel2014chiral}, a
dense region in the energy spectrum can be discerned. This is a
signature of enhanced degeneracy which in turn is a consequence of the
presence of two minima (maxima) in the lower (higher) energy band.
Beyond ${\xi_{d}}_{c}$, in the Meissner phase, energy density is
diminished. The critical point $ {\xi_{d}}_{c}$ is the point at which
the two minima merge into one, and can be analytically computed for
general $\varphi$:
\begin{equation}
\label{eqs:7}{\xi_{d}}_{c}=\begin{cases}\frac{-(\xi+\cos\varphi)+\sqrt{(\xi-\cos\varphi)^2+4\sin^2\varphi}}{2},  &\xi\leq\frac{\sin^2\varphi}{\cos\varphi}\\
\hspace{60pt}0, & \xi\geq\frac{\sin^2\varphi}{\cos\varphi}
\end{cases}
\end{equation}
Conversely, in terms of $\xi_d$ the critical point $\xi_c$ is given by
\begin{equation}
\label{eqs:8}{\xi}_{c}=\begin{cases}\frac{\sin^2\varphi-{\xi_d}^2-\xi_d\cos\varphi}{\xi_d+\cos\varphi},  &\xi_d\leq\frac{-\cos\varphi+\sqrt{1+3\sin^2\varphi}}{2}\\
\hspace{40pt}0, & \xi_d\geq\frac{-\cos\varphi+\sqrt{1+3\sin^2\varphi}}{2}
\end{cases}.
\end{equation}
Eq.(~\ref{eqs:8}) shows that the critical point $\xi_c$ decreases on
increasing diagonal hopping upto $\xi_d=1$ as already suggested by
a study of chiral current in Fig.~\ref{figs:2}(b). For $\xi_d \ge 1$, $\xi_c=0$ because the relative critical
value is never negative. 
A further look at Fig.~\ref{figs:4}(a) reveals
that, there is a reappearance of dense states in the topological
region. This is a consequence of degeneracy due to the appearance of two \emph{maxima} (\emph{minima})
in the lower (higher) energy band. This is a new phase phase transition, different from vortex-Meissner, and 
appears to have not been reported earlier and has features of a van Hove singularity~\cite{ashcroft1976introduction}. This new critical point is given by
\begin{equation}
\label{eqs:9}{\xi^{'}_{d}}_{c}={\Bigg(\frac{\xi+\sqrt{\xi^2+4}}{2}\Bigg)}_{\varphi=\frac{\pi}{2}}.
\end{equation}
We have computed this point, only for the case $\varphi = \frac{\pi}{2}$, where the bandstructure is symmetric (Fig.~\ref{figs:4}(a)).
The maxima of the lower band appear at the zone boundary when $\xi\le{\xi^{'}_d}_{c}$. But for $\xi_d >{\xi^{'}_{d}}_{c}$ they start to move inside resulting
in degeneracy of other eigenstates. The maximum possible shift is $k=\pi/2$, which corresponds to the 
limit $\xi_d\to\infty$ when the ladder model behaves as two independent nearest neighbour tight-binding chains.

\section{\label{sec:level1} Many body entanglement with vertical division}
\begin{figure}[h!]
\centering
\includegraphics[scale=0.85]{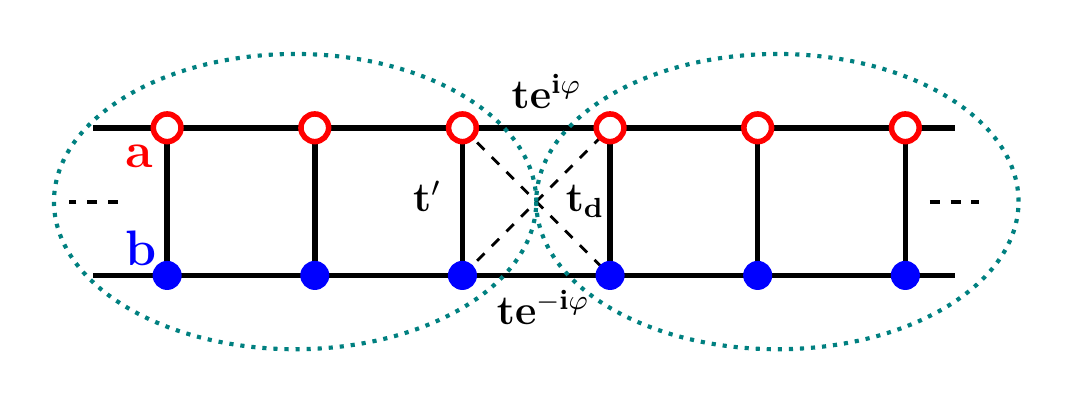}
\caption{Ladder system with vertical division of two subsystems.}
\label{figs:6}
\end{figure}
Another useful division for the purpose of studying entanglement
entropy is to consider two subsystems obtained by cutting the ladder
vertically as shown in Fig.~\ref{figs:6}.  In the vertical division
case, the entanglement entropy seems to be closely correlated with the
band spectrum.  For $\phi = \pi/2$, the band spectrum is symmetric
about the centre, as shown in Fig.~\ref{figs:4}(a). The band closes at
$\xi_d = \xi$ and forms two degenerate edge modes when the boundary
conditions are open.

\begin{figure}[h!]
\includegraphics[scale=0.34,angle=-90]{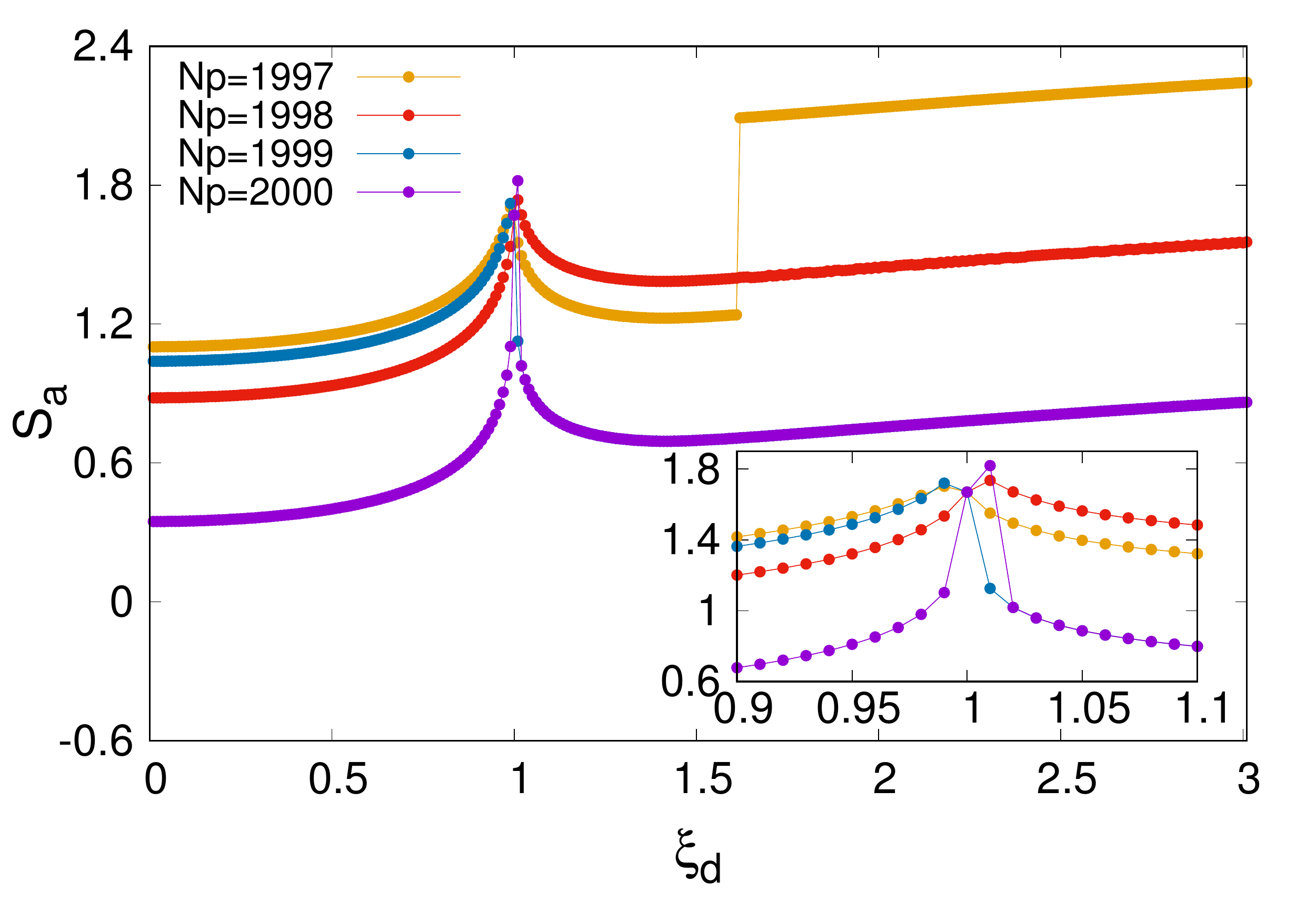}
\caption{Subsystem entanglement entropy ($S_a$) with $\xi_d$ for $L=1000$, $\xi=1$, $\varphi=\frac{\pi}{2}$ with open boundary conditions.}
\label{figs:7}
\end{figure}

\begin{figure}[h!]
\includegraphics[scale=0.34,angle=-90]{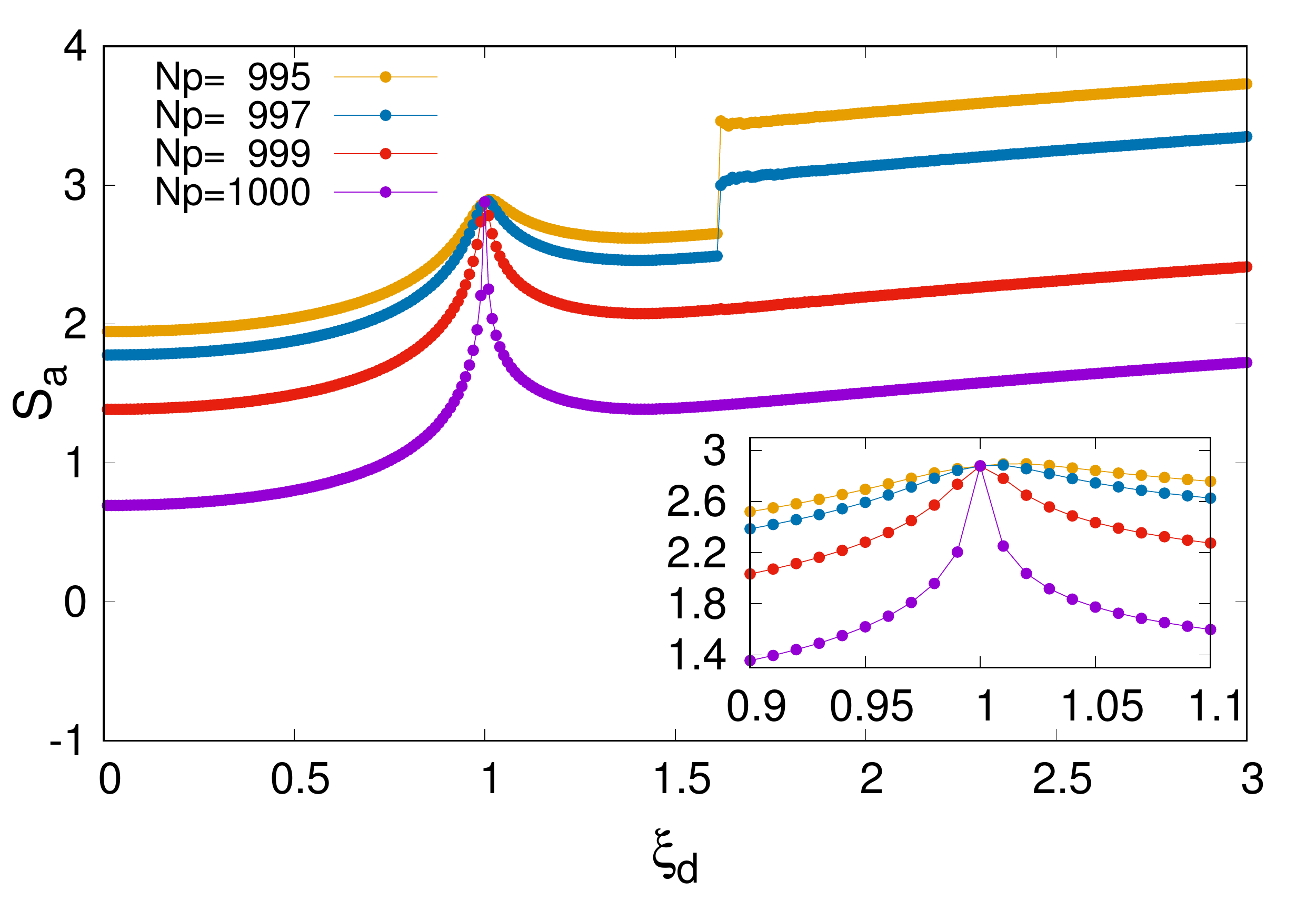}
\caption{Subsystem entanglement entropy ($S_a$) with $\xi_d$ for $L=1000$, $\xi=1$, $\varphi=\frac{\pi}{2}$ with periodic boundary conditions.}
\label{figs:8}
\end{figure}

When the boundary conditions are closed, the band gap closes at $\xi_d
=\xi$, but the gap is non-zero on either side of the topological phase
transition. From Fig.~\ref{figs:7} and Fig.~\ref{figs:8}, we observe
that at the topological phase transition point, the entanglement
entropy takes on the same identical value for a range of fillings
close to half-filling. Furthermore, with periodic boundary conditions,
the entanglement entropy is maximum at the topological phase
transition. The point where entanglement entropy attains a common
value for a variety of fillings close to half-filling, and around
which it splays out for different fillings, is thus a useful method to
capture a topological phase transition. The magnitude of the
entanglement here is much smaller than that obtained within horizontal
division, because here it is only one (or two in the case of periodic
boundary conditions) plaquette that contributes to the entanglement,
whereas with horizontal division, every rung contributes, thus making
the overall entanglement of the order of the number of particles
$N_p$.

A further feature that entanglement entropy is sensitive to, is the
enhanced degeneracy that appears in the band structure at the critical
point described by Eq.~\ref{eqs:9}. The sharp increase in density of
states might be indicative of a van Hove singularity. A jump in
entropy appears when the particle filling is half-filling minus three
as shown in Fig.~\ref{figs:7} and Fig.~\ref{figs:8} both of which have
this point at $\xi_d = 1.618$, in agreement with Eq.~\ref{eqs:9}.  The
reason for this goes back to the argument we employed to reach
Eq.~\ref{eqs:9}.  When particles are removed starting from
half-filling (which results in only one of the two edge states being
occupied), the first particle is removed from the edge state. The next
two particles then come off from the maxima in the bands, and then
there are four degenerate levels for the next particles. It is at this
filling level, with the enhanced degeneracy, that entanglement entropy shows
the sensitivity.
\begin{figure}[h!]
\includegraphics[scale=0.168,angle=-90]{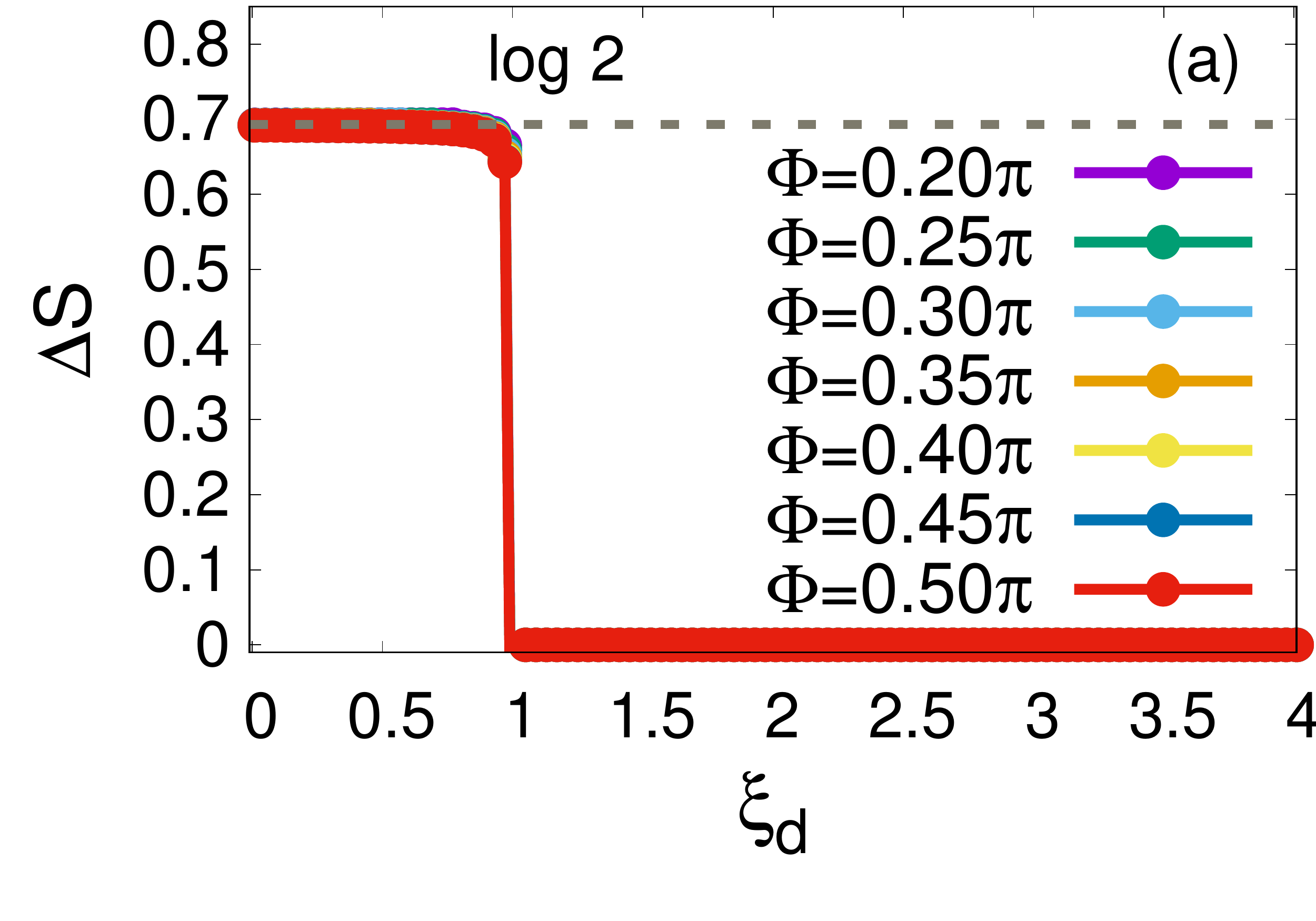}\
\includegraphics[scale=0.168,angle=-90]{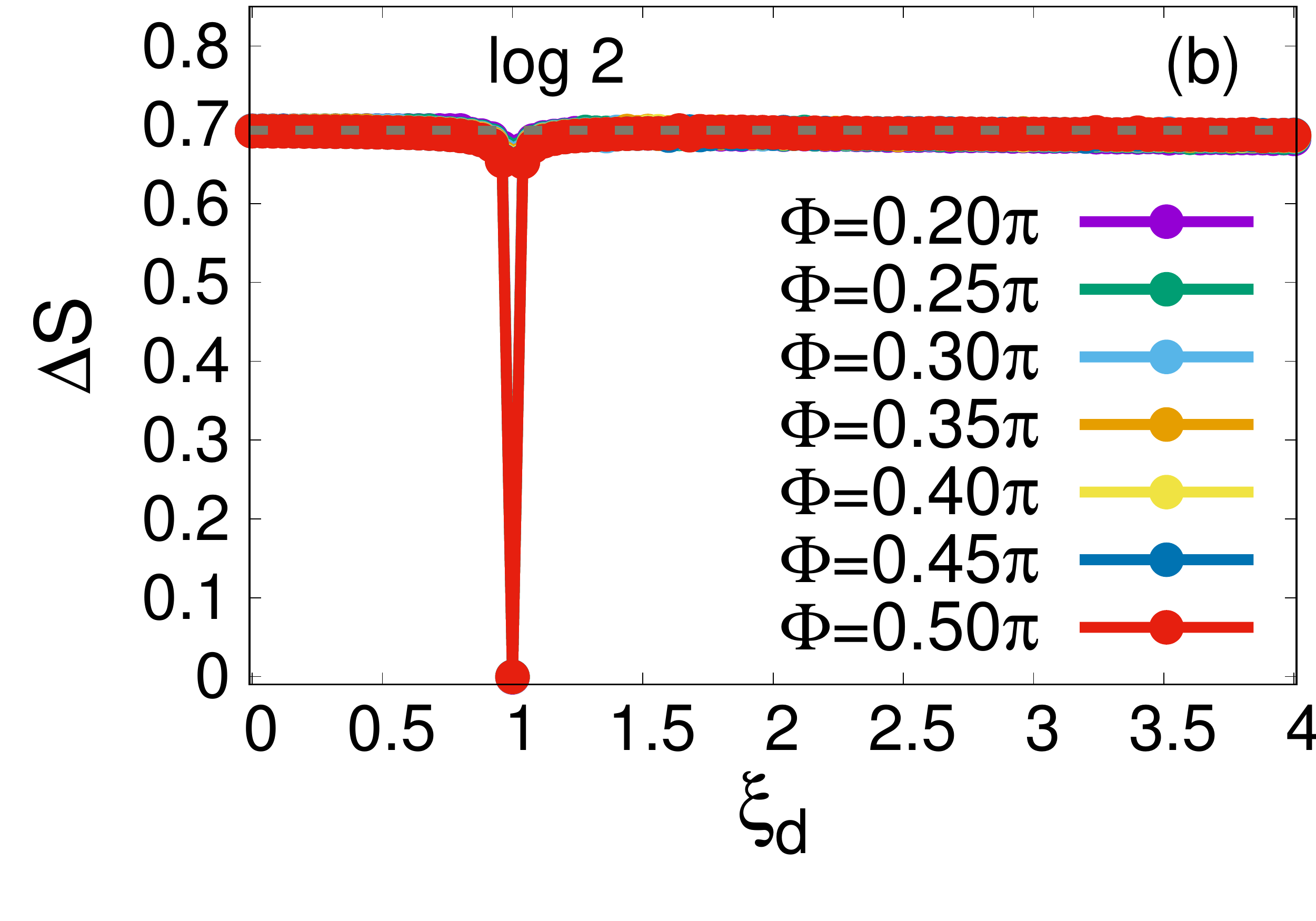}
\caption{Subsystem entanglement entropy of ladder system ($\Delta S_a$) with $\xi_d$. (a)OBC, (b) PBC.}
\label{figs:9}
\end{figure}
Once again, calculating the entanglement entropy difference between
half filled and one less than half filled captures the trivial to
topological phase transition, as shown in Fig.~\ref{figs:9}.  With
open boundary conditions, for $(\xi_d < \xi)$, there is a finite value
of $\Delta S = \log2$ whereas after $(\xi_d =\xi)$ due to the
degeneracy of the edge modes the entropy difference $\Delta S$ tends
to zero. However, in the case of periodic boundary conditions, the gap
is closed only at $\xi_d = 1$, hence the entropy difference goes to
zero at that point alone. Once again, we see that this is a feature of the topological phase transition, and is thus completely independent of $\varphi$.
\bibliography{supplementaryECL}